\documentclass[12pt]{iopart}

\usepackage{graphicx}
\begin{document}

\title[Electric and magnetic field effects on excitons in core-multishell quantum wires]{Electric and magnetic fields effects on the excitonic properties of elliptic core-multishell quantum wires}

\author{R. Mac\^edo$^{1}$, J. Costa e Silva$^{1}$, Andrey Chaves$^{2}$, G. A. Farias$^2$, and R. Ferreira$^{3}$}
\address{$^1$Departamento de Ci\^{e}ncias Exatas e Naturais,
Universidade Federal Rural do Semi-\'Arido, Caixa Postal 6030,
59600-900 Mossor\'o, RN, Brazil}
\address{$^2$Departamento de F\'isica, Universidade Federal do
Cear\'a, Caixa Postal 6030, Campus do Pici, 60455-900 Fortaleza,
Cear\'a, Brazil}
\address{$^3$Laboratoire Pierre Aigrain, Ecole Normale Superieure,
CNRS UMR 8551, Universit\'e P. et M. Curie, Universit\'e Paris
Diderot, 24 rue Lhomond, F-75005 Paris, France}
\ead{\mailto{andrey@fisica.ufc.br}}

\begin{abstract}
The effect of eccentricity distortions of core-multishell quantum wires on their electron, hole and exciton states is theoretically investigated. Within the effective
mass approximation, the Schr\"odinger equation is numerically
solved for electrons and holes in systems with single and double
radial heterostructures, and the exciton binding
energy is calculated by means of a variational approach. We show that the energy spectrum of a core-multishell heterostructure with eccentricity distortions, as well as its magnetic field dependence, are very sensitive to the direction of an externally applied electric field, an effect that can be used to
identify the eccentricity of the system. For a double
heterostructure, the eccentricities of the inner and outer shells
play an important role on the excitonic binding energy, specially in the presence of external magnetic fields, and lead to drastic modifications in the oscillator strength.
\end{abstract}

%Uncomment for PACS numbers title message
% Keywords required only for MST, PB, PMB, PM, JOA, JOB?
%\vspace{2pc}
%\noindent{\it Keywords}: Article preparation, IOP journals
% Uncomment for Submitted to journal title message
%\submitto{\JPA}
% Comment out if separate title page not required
\maketitle

\section{Introduction}

The investigation of semiconductor low-dimensional structures has
a key role on the development of nanotecnology and future
nanoscale devices. Recent advances in growth techniques made
possible the fabrication of quantum wire structures, where
electrons and holes are confined in two dimensions, with only one
free dimension. Quantum wires based on radial semiconductor
heterostructures have been successfully used in opto-electronic
systems and biological sensors, which require a strong confinement
of the charge carriers \cite{D. Appell,Y. Cui,M. Law}. Many
research groups have used various growth techniques to synthesize
quantum wires composed of two different semiconductor materials,
forming several types of wire, for example i) core-shell wires,
\textit{i.e.} a strand of a semiconductor material covered by a
layer of another semiconductor material, ii) core-multishell
structures, where a wire is covered by a sequence of layers of
other materials, and iii) superlattice nanowires, where a
longitudinal superlattice is built by stacking cylindrical blocks
of different semiconductor materials \cite{H.-J. Choi,J.
Goldberger,Y. Wu, Lauhon}, just to mention a few. Photoluminescence experiments on
GaAs/AlGaAs \cite{Kang, Fickenscher} and InAs/InP \cite{Zanolli,
Mohan} core-shell quantum wires have been recently reported, where
quantum confinement of the carriers is observed with energies that
are in good agreement with theoretical predictions.

In this paper, we present a theoretical study of the excitonic
properties of core-multishell (CMS) quantum wires, focusing on the
effects of eccentricity distortions of the cylindrical wire
structure on the exciton binding energy and on its behavior under
external electric and magnetic fields. The main effect of the
eccentricity is to create regions of higher curvature and,
consequently, lower effective potential, \cite{Raimundo, Gridin}
which work as traps to the otherwise circularly symmetric charge
carriers wavefunctions. A similar effect is investigated
\textit{e.g.} in Ref. \cite{Ferrari}, though only for
electrons, not for excitons, in the so-called prismatic quantum
wire, where a hexagonal-like wire exhibits a higher wavefunction
distribution over the edges of the hexagon, where the curvature is
higher. In this case, the electron energy states exhibit
oscillations as the intensity of an axially applied magnetic field
increases, which is reminiscent of the Aharonov-Bohm (AB) effect.
\cite{Ferrari} In fact, the AB effect for electrons in a
core-shell wire has been experimentally confirmed by the
observation of quantum interference effects on the
magnetoresistance of a In$_2$O$_{3}$/InO$_x$ nanowire. \cite{Jung} 

As the exciton has neutral net charge, it is not supposed to be
affected by electric/magnetic fields \textit{a priori}. \cite{Song}
However, electric fields naturally separate electrons and holes
wavefunctions, due to their opposite charges, creating regions of
non-neutral net charge and affecting the exciton binding energy, 
\cite{Maslov} through the so-called quantum confined Stark effect (QCSE). Our results show how the carriers energy spectra and
exciton binding energy are affected by an applied electric field
perpendicular to the wire axis, considering different in-plane
directions, which can be used to obtain information about the
eccentricity of the wire and its direction of distortion. Such a
charge carriers separation does not occur in the case of an
axially applied magnetic field, leading to a practically neutral
net charge and, consequently, a negligible effect of the field
intensity on the exciton binding energy. On the other hand, the
observation of excitonic AB oscillations have been reported in
semiconductor quantum \textit{rings} when electrons and holes have
an intrinsic spatial separation, due to built-in electric fields
or structural effects. \cite{excitonrings, Santander, Teodoro,
BinLi, 4} In fact, our results demonstrate that in elliptic CMS
wires, electrons and holes are pushed to the regions of higher
curvature of the confining shell with different strengths,
yielding regions of non-neutral charge and, consequently, to an
enhancement of the magnetic field effect on the exciton binding
energy, which starts to exhibit AB oscillations. The regions of non-neutral charge demonstrated in this paper may also play a role in other geometries of core-shell quantum wires which share the property of exhibiting high curvature regions, such as the semiconductor quantum wires with square, hexagonal and triangular cross sections reported in the literature. \cite{Fortuna, Ferrari}

The remainder of the paper is organized as follows: Sec. II
presents the theoretical model for calculating the exciton energy.
The results for quantum wires containing one or two
confining shells are shown and discussed in Sec. III, whereas in
Sec. IV we summarize our findings and present our concluding
remarks.

\section{Theoretical Model}
\label{sec:model}

The system we investigate consists of a cylindrical inner wire
(core) covered by a sequence of layers of different intercalated
materials (shells), as sketched in Fig. 1 (a). In order to help us
to study non-circular symmetries of the system, Cartesian
coordinates are applied, taking the $(x,y)$ as the confinement
plane and $z$ as the free direction. The eccentricity of the
elliptic core-multishell structure, which is top viewed in Fig. 1
(b), is defined as $\xi = a/b$. The ellipse is assumed to preserve
the same inner area as the original circular shell that generates
it. Using the symmetric gauge for the vector potential,
$\textbf{A} = \left(By/2, -Bx/2, 0\right)$, the Hamiltonian that
describes the system, within the effective mass approximation,
reads \cite{F. M. Peeters, CostaeSilvaQWR}
\begin{eqnarray}
H_{exc} = \sum_{i = e, h} \left\{- \frac{\hbar^2}{2 m_i^{\parallel}}\left(\vec{\nabla}_{2D,i} - \frac{2\pi}{\Phi_0}\vec A\right)^2
+ V_i(x_i, y_i)\right\} \nonumber \\
- \frac{\hbar^2}{2\mu_{\perp}} \frac{\partial^2}{\partial z^2}  - \frac{e^2}{\epsilon\vline \vec{r}_e - \vec{r}_h \vline}
 \label{eq2.1},
\end{eqnarray}
where the indexes $e$ and $h$ stand for electron and hole,
respectively, $m_i^{\parallel}$ is the in plane effective mass of each charge
carrier, $\Phi_0 = h/e$ is the magnetic quantum flux, $\mu_{\perp}$ is the electron-hole reduced mass in the $z$-direction, and $z = \mid z_e - z_h \mid $ is the
electron-hole relative coordinate. The $V_{i}(x_i,y_i)$ term is the hetero-structure potential, due to the bands
mismatches, whereas the last term of Eq. (\ref{eq2.1}) accounts for the electron-hole Coulomb interaction.

\begin{figure}[h!]\label{fig:sketch}
\centerline{\includegraphics[width=0.7\linewidth]{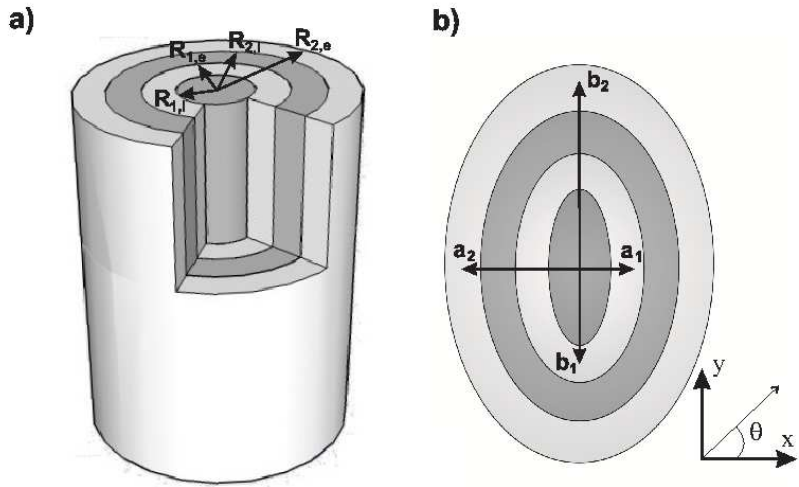}}
\caption{(a) Sketch of a core-multi-shell structure studied in
this work, which is infinitely long in the $z$-direction and has a
radial heterostructure made out of different semiconducting
materials, represented by light and dark grey regions. The $m$-th
shell is delimited by inner and outer radii $R_{m,i}$ and
$R_{m,e}$. We also investigate elliptic structures, where the
eccentricity is defined as $\xi = a_m/b_m$, as illustrated in the
top view of the elliptic core-multishell quantum wire in (b).}
\end{figure}

The influence of a graded interface between materials has been
already investigated by previous papers in the literature,
\cite{CostaeSilvaQWR} where it was demonstrated that such an
interface just leads to slightly larger confinement and excitonic
energies, while their dependencies on external magnetic fields or wire dimensions
remains essentially preserved. In this work, we will consider an abrupt
interface between the core and shell layers. Dielectric mismatch effects between materials are also neglected in our model. Previous papers in the literature \cite{BinLi2,Slachmuylders,Ariel} have demonstrated that these effects are described by potentials that are proportional to $1-\epsilon_1/\epsilon_2$, where $\epsilon_1$ and $\epsilon_2$ are the dielectric constants of the hetero-structure materials, and are less important for large widths of the hetero-structure layers. The materials involved in the hetero-structures studied in this paper have aproximately the same dielectric constant, and the role of the dielectric mismatch effect due to the boundary between the outmost shell and the surrounding medium is ruled out by considering a large width of this shell, therefore, dielectric mismatch effects are negligible to a good approximation in our model. The heterostructure materials considered in this work have also similar lattice constants, so that strain effects does not play an important role as well.

In order to solve the Schr\"odinger equation for the exciton $H_{exc}\Psi = E_{exc}\Psi$, we neglect initially the coulombic coupling, so that the solutions can
be put in the form $\Psi = \psi_e(\vec \rho_e)\psi_h(\vec \rho_h)\psi_z(z)$ to obtain
\begin{equation}\label{eq.2DSchrodinger}
\left\{- \frac{1}{m_i^{\parallel}}\left(\vec{\nabla}_{2D,i} - \frac{2\pi}{\Phi_0}\vec A\right)^2
+ V_i(x_i, y_i)\right\}\psi_i = E_i \psi_i,
\end{equation}
for each carrier $i$, whose in-plane position vector is given by
$\vec{\rho_i}$. In this equation, and from now onwards, energy
variables are divided by the Rydberg energy $Ry$, spatial
variables by the Bohr radius $a_0$, and effective masses by the
free electron mass $m_0$, in order to make the equations
dimensionless. A gauge invariant form of the finite differences
scheme is applied to solve for the lowest energy states of this equation numerically,
\cite{Governale} from where one obtains the carriers in-plane
wavefunctions $\psi_e$ and $\psi_h$, which are then integrated
along with the Coulomb term to form an effective potential for the
problem in the $z$-direction, yielding
\begin{eqnarray}\label{eq.SchrodingerZ}
-\frac{1}{\mu_{\perp}} \frac{d^2}{d z^2}\psi_z - \frac{2}{\epsilon}\int d\vec{\rho_e}\int d\vec{\rho_h}\frac{\vline\psi_e\vline^2\vline\psi_h\vline^2}{\sqrt{ (\vec{\rho}_e - \vec{\rho}_h)^2 + z^2}}\psi_z \nonumber \\ = E_z \psi_z.
\end{eqnarray}

We assume a gaussian wave function \cite{Frank L.
Masdarasz} as the solution in $z$
\begin{equation}\label{eq.Gaussian}
\psi_z(z) = \frac{1}{{\sqrt \eta  }}\left( {\frac{2}{\pi }}
\right)^{1/4} \exp \left( { - \frac{{z^2 }}{{\eta ^2 }}} \right),
\end{equation}
where $\eta$ is the variational parameter that minimizes $E_z =
\langle \psi_z| H_z | \psi_z\rangle$, where $H_z$ is the
Hamiltonian in the $z$-direction (see Eq.
(\ref{eq.SchrodingerZ})). Some of the integrals involved in this
minimization are analytical, \cite{CostaeSilvaQWR}
leading to
\begin{equation}
E_z  = \frac{1}{{\mu_\bot\eta ^2}} - \frac{2}{\varepsilon \eta}\sqrt {\frac{2}{\pi }} \int\limits_V{\left| {\psi _e} \right|^2} {\left| {\psi_h} \right|^2} \exp \left(a \right) K_0 \left(a \right)dV,
 \label{eq2.11}
\end{equation}
where $dV = d\vec{\rho_e}d\vec{\rho_h}$, $a = \left| \overrightarrow {\rho_e} - \overrightarrow{\rho_h} \right|^2/\eta^2$ and $K_0(x)$ is
the modified zero-order
Bessel function of second kind. The last integral is further calculated numerically, and the exciton energy is obtained by $E_{exc}= E_e + E_h - E_z$. We repeat the variational procedure for different combinations of electron and hole eigenstates and look for the combination of states that lead to the lowest exciton energy $E_{exc}$, which, as far as the variational method is concerned, represents an upper bound for the exciton ground state energy. 

Notice that, rigorously speaking, the best approach to this problem would be not to separate the electron and hole variables, since Eq. (\ref{eq2.1}) is actually non-separable, because of the Coulomb term. However, such an exact approach would lead to a five-dimensional problem (two in-plane coordinates for each electron and hole and a relative $z$-coordinate for the exciton), which is computationally expensive. Using the electron-hole states basis to construct a matrix that represents the Hamiltonian and then diagonalizing it is not convenient as well, since each matrix element would consist of a five-dimensional integral, and there must be many of them for a properly accurate Hamiltonian. As a matter of fact, these more accurate procedures were used in the literature mostly for circularly symmetric problems, where one can take advantage of the system symmetry to reduce the number of coordinates, or to (semi-)analytically solve for the single-particle eigenstates. \cite{Song, Mehran} Conversely, our case requires Cartesian coordinates, which allows us to investigate elliptic shells, in-plane electric fields, and systems with arbitrary shell geometry, by paying the price of having more coordinates to deal with. Thus, the main idea of the present model is to circumvent this problem by using a much simpler approach, which, in summary, consists in (i) assuming, as an approximation, that one can separate the variables, (ii) solving for the single-particles separately, and then (iii) minimizing the exciton energy as a whole by a variational procedure. Due to its simplicity, this factorization of the wave function has been widely used in the literature, even in the case of degenerate single particle states, such as in AB oscillations. \cite{Janssens} However, it is worthy to point out that in the degenerate case, this procedure leads to less accurate results, since it does not allow for an excitonic state consisting of an admixture of single-particle eigenstates. Even so, we believe that, despite the model simplicity, the results obtained by the present approximation in the remainder of this paper are accurate enough, and a more sophisticated technique would require great efforts for just small quantitative corrections to the results. 

\section{Results and Discussions}

In the following subsections, we apply our model for the study of the eigenstates and binding energies of systems consisting of a Al$_{0.3}$Ga$_{0.7}$As core wire covered by one and
two confining GaAs shells, respectively. In our calculations, we assume the dielectric constant of GaAs as $\epsilon$ = 12.9 $\epsilon_0$, the conduction (valence) band offset $V_e$ = 262 meV ($V_{hh}$ 195 meV), the electron effective mass (assumed to be isotropic) as $m_e^{\bot} = m_e^{\|} =$ 0.067 $m_0$, and the heavy hole effective mass as $m_{hh}^{\bot} = 0.11 m_0$  and $m_{hh}^{\|} = 0.51 m_0$. \cite{parameters2, parameters3}

\subsection{Single shell confinement}

Let us first consider a Al$_{0.3}$Ga$_{0.7}$As core wire, covered
by a GaAs/Al$_{0.3}$Ga$_{0.7}$As shell. In this case, the GaAs shell works as a radial quantum well for both electron and hole, which will be
preferably confined in this region.

\begin{figure}[h!]
\centerline{\includegraphics[width=0.7\linewidth]{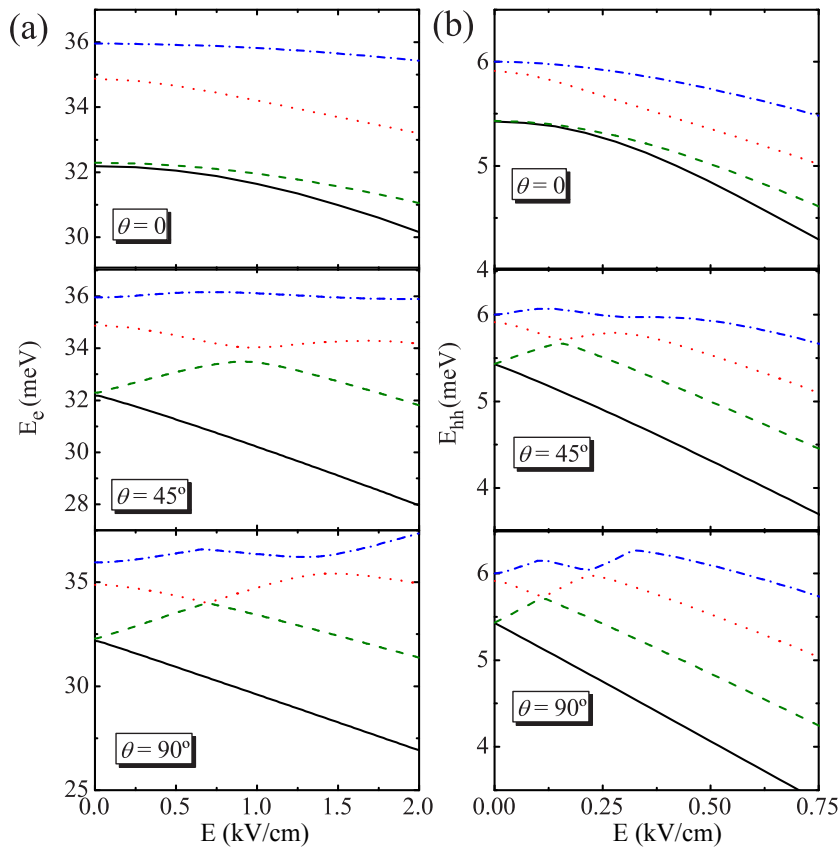}}
\caption{(color online) Confinement energies of (a) electrons and (b) holes in a
Al$_{0.3}$Ga$_{0.7}$As/GaAs/Al$_{0.3}$Ga$_{0.7}$As radial
heterostructure as a function of the in-plane electric field
intensity, considering an eccentricity $\xi = 0.9$. The field is
applied at three different angles $\theta$ from the
$x$-axis.}\label{fig:ellipsingleshell}
\end{figure}

\begin{figure}[!h]
\centerline{\includegraphics[width=0.4\linewidth]{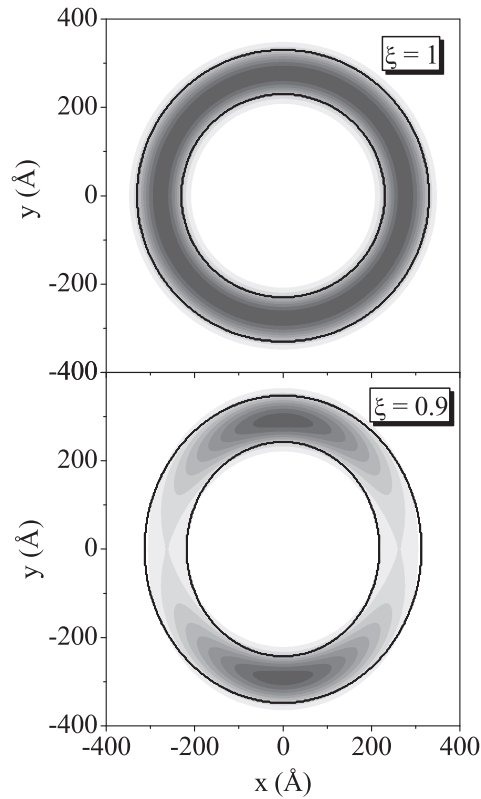}}
\caption{Electron ground state wave functions in (a) circular and (b) $\xi = 0.9$ elliptic Al$_{0.3}$Ga$_{0.7}$As/GaAs/Al$_{0.3}$Ga$_{0.7}$As core-multishell
 quantum wires, considering R$_{1,i}$ = 230 \AA\, and R$_{1,e}$ = 330 \AA\,. Similar localization effects are obtained for holes (not shown).} \label{fig:ellipsingleshellPSI}
\end{figure}

Let us focus initially on the effects of an external electric field applied perpendicularly to the wire axis of a core wire on its energy states. Obviously, in a circular heterostructure, the energy spectrum does not depend on the electric field application direction in the $(x,y)$-plane. Moreover, as we previously mentioned, the
confined states in the circular case are similar to the in-plane
states of circular quantum rings, where the influence of an
in-plane electric field has already been extensively investigated
by previous works in the literature. \cite{Latge1, Maslov}
In the elliptic CMS
case, electric fields applied in different directions lead
to qualitatively different behaviors of the energy states as a
function of the electric field modulus, as illustrated in Fig.
\ref{fig:ellipsingleshell}, for the (a) electron and (b) heavy
hole energy states in a $\xi = 0.9$ elliptic CMS, considering an
electric field with different application angles $\theta$ (we consider R$_{1,i}$ = 230 \AA\, and R$_{1,e}$ = 330 \AA\,). More
specifically, for a wire squeezed in the $x$-direction (see Fig.
1), when the electric field is applied in the $y$-direction
($\theta = \pi/2$), the ground state rapidly decreases with the field (by roughly $eF(R_{1,e}+R_{1,i})/2$), while many crossings are observed between excited
states as the field intensity increases, whereas for a field
applied in the squeezing direction ($\theta = 0$), the energies
just decrease with the field, due to the quantum confined stark effect. \cite{AndreyWell} Moreover, the energy crossings for
electrons and holes happen for different values of electric field,
which could help one to identify which carrier is involved in each
transition. Actually, the behavior observed in the $\theta =
\pi/4$ case is quite similar to the one in stacked double quantum
dots under an electric field applied along the dots axis.
\cite{Bastard, QDField} This is a consequence of the localization
of the carriers in the regions of maximum curvature, as
illustrated in Fig. \ref{fig:ellipsingleshellPSI}, so that when
the electric field is applied at $\theta = \pi/4$, the system is
indeed similar to the double dots case mentioned. So the dependence of the energy levels with both magnitude and orientation of the applied electric field can be used as a tool to infer on the eccentricity of
experimentally grown CMS wires, where the presence of crossings
between energy states indicates in which direction the higher
curvature regions are situated.

\begin{figure}[!h]
\centerline{\includegraphics[width=0.6\linewidth]{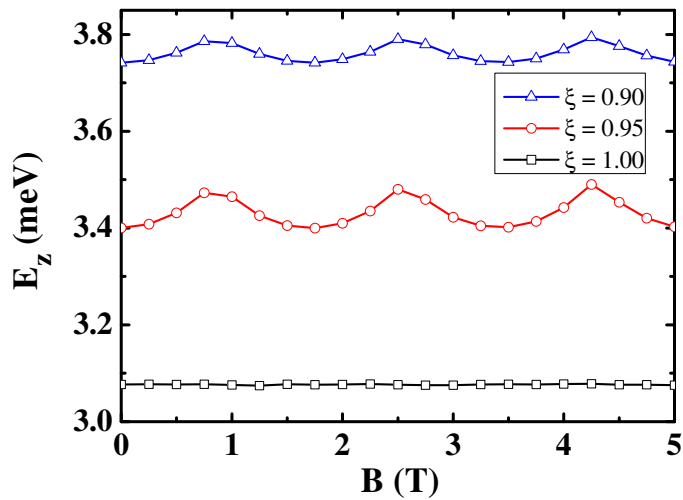}}
\caption{(color online) Exciton binding energies in a Al$_{0.3}$Ga$_{0.7}$As/GaAs/Al$_{0.3}$Ga$_{0.7}$As radial heterostructure
as a function of the magnetic field for different values of eccentricity: $\xi = 1$ (black squares), 0.95 (red circles) and 0.9 (blue triangles).} \label{fig:BindingSingleKSI}
\end{figure}

By analyzing Fig. \ref{fig:ellipsingleshellPSI}, it becomes clear
that the in-plane wave functions in this system are quite similar
to those in quantum rings, as expected. \cite{CostaeSilvaRings,
CostaeSilvaSSC, Gridin} The difference between the two systems concerns the wave functions in $z$-direction, which are spatially restricted in the latter, but
free in the former. As in the rings case, in the presence of an
axially applied magnetic field, the
eigenenergies of individual carriers display oscillations due to the
Aharonov-Bohm effect  (not shown, see e.g. Refs. \cite{CostaeSilvaRings, Wendler}). On the other hand,
the exciton is a neutral particle and, as such, should not exhibit
response to magnetic fields. Indeed, as one can verify in the Fig.
\ref{fig:BindingSingleKSI} (solid curve), there is no significant
modification of the exciton binding energy as the magnetic field
increases for a circular shell. Interestingly enough, the AB
effect is significantly enhanced as the eccentricity of the CMS
becomes lower than 1. This effect comes from the fact that electron and hole wave
functions are not completely overlapped in the elliptic cases
shown: the integrated overlap between electron hole for each case
were calculated as $\langle \psi_e | \psi_h \rangle \approx$
0.999, 0.982 and 0.984, for $\xi = 1$, 0.95 and 0.9, respectively.
In the circular case, there is no preferable confinement region
along the shell, so that both the electron and hole are
distributed equally along the whole GaAs region and the net charge
is neutral practically everywhere. The integrated overlap does not
reach unity in this case just because their wavefunctions penetration in the barriers are slightly different.
However, for a $\xi = 0.95$ elliptic system, the carriers are
pushed to the highest curvature regions, and in the proposed CMS,
heavy holes are pushed more strongly, due to their heavier mass as
compared to that of the electron. Being still not completely
confined in the high curvature regions, the electron wavefunction
covers a larger region as compared to the hole's wave function,
then creating a wide region of suppressed overlap, namely, a
non-neutral net charge. As a consequence, the AB effect is
enhanced, as shown by the curve with circles in Fig.
\ref{fig:BindingSingleKSI}. As the eccentricity value is further
decreased to $\xi = 0.90$, the electron is further compressed
towards the high curvature regions and, eventually, electron and
hole wavefunctions start to cover similar areas, enhancing again
the integrated overlap to $\langle \psi_e | \psi_h \rangle
\approx$ 0.984. Most importantly, being pushed towards the high curvature regions, the wave functions start loosing the ring-like topology of the shell and, consequently, the amplitude of oscillation in the $\xi = 0.95$ case is slightly higher than that of $\xi = 0.9$. Such a reduction in the amplitude of AB oscillations due to the eccentricity of a ring confinement is also observed in the single particle energy spectrum of quantum rings, as reported in Ref. \cite{CostaeSilvaRings}.

\begin{figure}[!h]
\centerline{\includegraphics[width=0.6\linewidth]{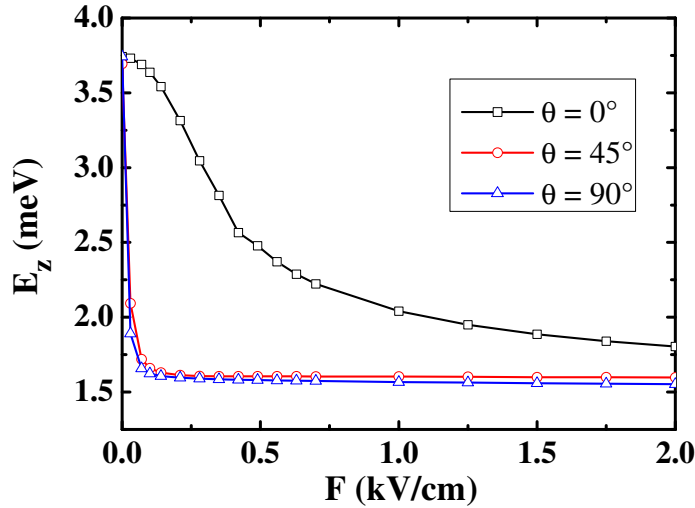}}
\caption{(color online) Exciton binding energies in a Al$_{0.3}$Ga$_{0.7}$As/GaAs/Al$_{0.3}$Ga$_{0.7}$As radial
elliptic heterostructure, with $\xi = 0.90$ eccentricity, as a function of an in-plane electric field with
different application angles $\theta$, where $\theta$ = 0 is the horizontal direction in Fig. 1 (b).} \label{fig:BindingSingleF}
\end{figure}

The ground state exciton energy as a function of the electric
field intensity is shown in Fig. \ref{fig:BindingSingleF} for a
$\xi$ = 0.90 elliptic quantum ring, considering the same angles of
application as in Fig. \ref{fig:ellipsingleshell}. The zero field binding energy is around 3.74 meV. As the electric field intensity increases, electron and hole are pushed
to different directions, leading to a smaller modulus of the
exciton binding energy $E_z$. For a horizontally applied electric
field ($\theta = 0$), the exciton binding energy decreases
smoothly as the electric field intensity is enhanced. Conversely,
for the other application angles, the exciton binding energy decreases
abruptly, rapidly converging to $\approx$ 1.55 meV. Indeed, for $ F \parallel x$ carriers should leave the $\theta=\pm\pi/2$ regions and migrate to the $\theta = 0$ (hole) or $\theta = \pi$ (electrons) regions under the action of the field, whereas
carriers are already preferably localized along the field in the $F \parallel y$
case.

\subsection{Double shell confinement}

\begin{figure}[!h]
\centerline{\includegraphics[width=0.7\linewidth]{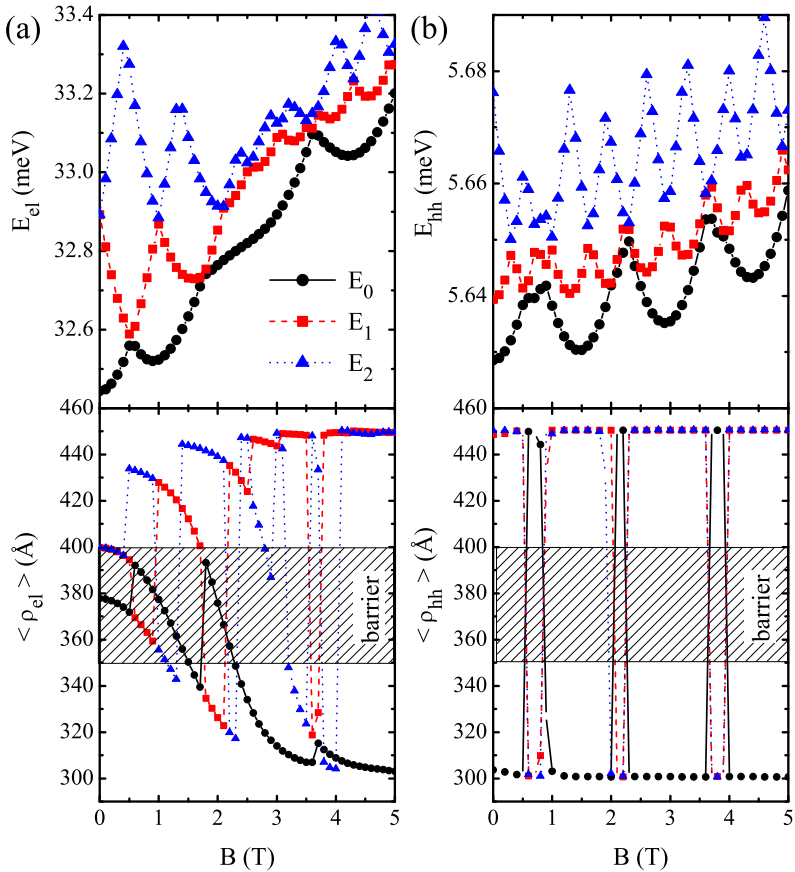}}
\caption{(color online) Confinement energies (top) and average radii (bottom) for
the low lying states of (a) electrons and (b) holes in a
Al$_{0.3}$Ga$_{0.7}$As/GaAs double shell structure, consisting of
a $R_{1,i}$ = 250 \AA\, and $R_{1,e}$ = 350 \AA\, inner shell and
a $R_{2,i}$ = 400 \AA\, and $R_{2,e}$ = 500 \AA\, outer shell. The
$W = $50 \AA\, Al$_{0.3}$Ga$_{0.7}$As barrier shell separating the
two GaAs confining shells is represented by the shaded area in the
bottom panels. Symbols represent the numerical results, and curves
are just guides for the eyes.} \label{fig:Energ2Cascas}
\end{figure}

We now investigate the case of a double shell structure,
consisting of an inner Al$_{0.3}$Ga$_{0.7}$As wire and two
confining (GaAs) shells with the same width $W = 100$ \AA\,,
separated by a 50 \AA\, barrier (Al$_{0.3}$Ga$_{0.7}$As) shell. Let us discuss initially the independent electron and hole states. In the circular case, the radial confinement energies for both
shells, separately, are practically the same, exhibiting only a
negligible difference due to the angular motion energy term, which
is inversely proportional the average radius of each shell. Figure \ref{fig:Energ2Cascas}
shows the confinement energies and average radii for (a) electrons
and (b) heavy holes states in such a structure, as a function of
the magnetic field intensity. The hole ground state (solid, black
circles) is initially confined in the inner GaAs shell, whereas
first (dashed, red squares) and second (dotted, blue triangles)
excited states are in the outer one, as one can verify by their
average radii in the bottom panel of Fig. \ref{fig:Energ2Cascas}
(b). This explains the larger period of oscillation of the hole
ground state, as compared to the excited ones. As the magnetic
field increases, the ground state curve reaches the other curves
periodically, and in these degeneracy points, the average radii of
these states are switched. For electrons, the situation is
slightly different: due to its lighter effective mass, the
electron states wave functions spread over both shells for lower
magnetic field intensities, so that their average radii end up in
the barrier region, as one can see in the bottom panel of Fig.
\ref{fig:Energ2Cascas} (a). However, for $B$ grater than $\approx$
2 T, the ground state stays in the inner shell region, whereas the
excited states are pushed to the outer confining shell, explaining
their smaller AB period, as compared to the one of the ground
state for larger fields, similar to the holes case.

\begin{figure}[!h]
\centerline{\includegraphics[width=0.6\linewidth]{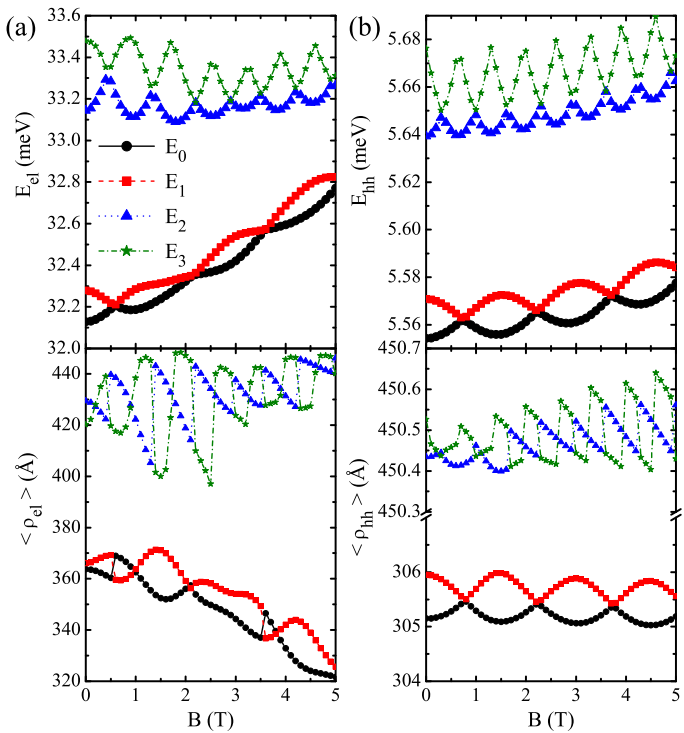}}
\caption{(color online) The same as in Fig. \ref{fig:Energ2Cascas}, but for a $\xi = 0.95$ elliptic inner shell.} \label{fig:Energ2CascasKsi}
\end{figure}

We previously demonstrated that the eccentricity pushes the charge
carriers towards the regions of largest curvature of the shell.
Hence, when one of the shells is elliptic, it creates an energy
imbalance between the inner and outer shells, creating a
preferable region of confinement and, consequently, avoiding
states that spread over both shells. This effect is expected to
smooth out the average radii transitions and the irregularity of
the AB oscillations period observed in Fig. \ref{fig:Energ2Cascas}
for the circular case. This is indeed observed in the results
shown in Fig. \ref{fig:Energ2CascasKsi} for the same structure as
in Fig. \ref{fig:Energ2Cascas}, but considering a $\xi = 0.95$
elliptic inner shell. In this case, for any value of magnetic
field intensity, ground and first excited
states are confined only in the inner shell,
whereas second and third (dashed-dotted, green stars) excited states
are in the outer one, which can be inferred by the different
periods of AB oscillations (top panels), as well as by their
average radii (bottom panels).

Actually, the analogous problem of two concentric quantum rings
was already investigated in a previous work,
\cite{CostaeSilvaRings} though it discusses only the electron
states, without mentioning excitonic properties. The carriers
eigenstates in the double shell quantum wire are quite similar to
those in the double rings case of Ref.
\cite{CostaeSilvaSSC}, which we suggest as a complementary
reading for more details about this topic.

\begin{figure}[!h]
\centerline{\includegraphics[width=0.6\linewidth]{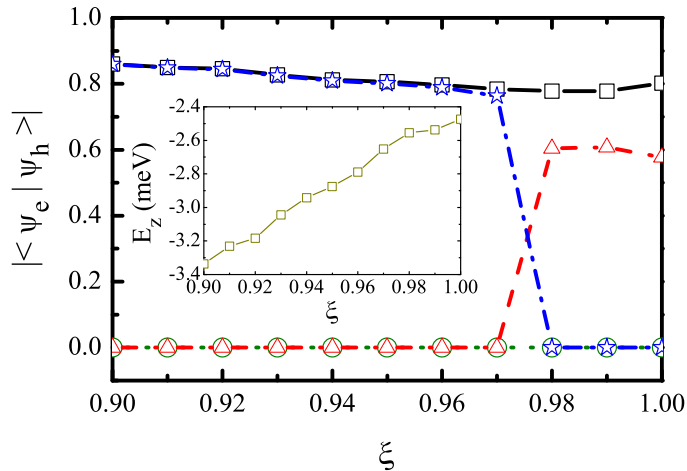}}
\caption{(color online) Overlap between electron and hole wave functions as a
function of the inner ring eccentricity, for $B = F = 0$. Overlaps are calculated
between the $n$-th electron state and the $m$-th hole state,
leading to the four ($n$,$m$) curves shown, where (0, 0) is
represented by squares, (0, 1) by triangles, (1, 0) by circles,
and (1, 1) by stars. The straight lines connecting the symbols are
guides for the eye. The inset shows ground state exciton binding
energies for these eccentricities.} \label{fig:OverlapsDouble}
\end{figure}

The ($n$,$m$) overlaps between the wave functions of the $n$-th
electron and the $m$-th hole eigenstate are shown as a function of
the inner shell eccentricity in the Fig. \ref{fig:OverlapsDouble},
where the ground state exciton binding energy is shown as inset (same CMS as in Figs. \ref{fig:Energ2Cascas} and \ref{fig:Energ2CascasKsi}, but for $B$ = $F$ = 0).
As we already mentioned, when both shells are circular, electrons
and holes spread over both shells in different ways, due to their
different masses and band-offsets. Therefore, in this case, the
overlap is small and, consequently, the binding energy of the ground state $(0, 0)$ is
depreciated. The modulus of the binding energy is enhanced as the
inner shell becomes elliptic, followed by the increase of the
electron-hole ground ground state state overlap (black squares).
Interestingly enough, the (0,1) overlap, between the ground state
of the electron and the first excited state of the hole (red
triangles), is non-zero only for $\xi > 0.96$, suggesting that
electron-hole transitions between these states are allowed only
for higher values of eccentricity. Moreover, the (1,0) overlap,
between the electron first excited state and the hole ground state
(green circles) is zero for any value of $\xi$, indicating that
the electron and hole first excited states are essentially
different.

\begin{figure}[!h]
\centerline{\includegraphics[width=0.7\linewidth]{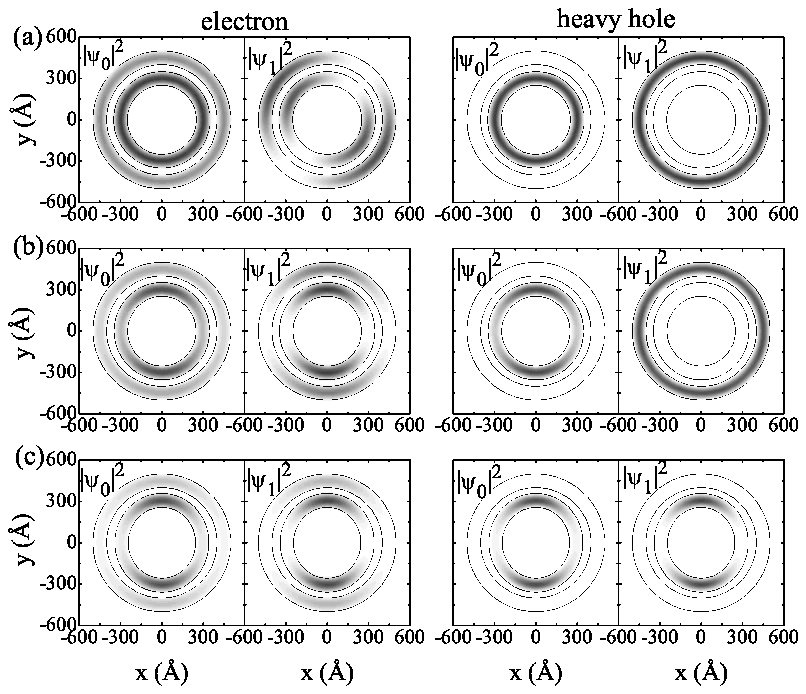}}
\caption{(color online) Electron (left panels) and heavy hole (right panels)
wavefunctions for a $R_{1,i}$ = 250 \AA\,, $R_{1,e}$ = 350 \AA\,,
$R_{2,i}$ = 400 \AA\, and $R_{2,e}$ = 500 \AA\,
Al$_{0.3}$Ga$_{0.7}$As/GaAs double shell structure, considering
different values of the inner ring eccentricity $\xi$ = 1.0 (a),
0.98 (b) and 0.95 (c), in the absence of both electric and magnetic fields.} \label{fig:2cascaswaves}
\end{figure}

The electron (left panels) and heavy hole (right panels) wave
functions shown in Fig. \ref{fig:2cascaswaves}, for different
values of eccentricity, help us to understand all the features
observed in Fig. \ref{fig:OverlapsDouble}. In the circular case
($\xi = 1$), the ground and first excited states of the hole are of $l_h = 0$ symmetry and
related to eigenstates of the radial confinement, localized at the
inner and outer shells, respectively. This is analogous to the
eigenstates of a double well structure with a large separation
barrier, where the ground and first excited states are related to
wave functions confined either in one or other of the two wells.
On the other hand, due to its smaller effective mass, the electron
tunnels more easily between inner and outer shells, The ground $l_e = 0$ states of the two independent shells hybridize to form roughly radially symmetric and anti-symmetric states.  The same holds for the $|l_e| = 1$ states, for which the tunnel coupling is greater, so that, in the end, the ground state of the whole structure ($n = 0$) is the $l_e = 0$ symmetric state while the first excited level ($n = 1$) is the $|l_e|=1$ symmetric one. Thus, for $\xi =
1$, (i) the (0,0) overlap is non-zero, but it is also lower than
1, because the hole state is entirely confined in the inner shell,
whereas the electron state spreads on both shells; (ii) the (0,1)
overlap is non-zero, since both states have the same angular
symmetry, and differ only by their charge distributions over the
inner and outer shells; and (iii) the (1,0) and (1,1) overlaps are
zero, since the first excited state of the electron has a
different angular symmetry, as compared to both states of the
hole. Notice also that the $s$-like ground state and the $p-$like
first excited electron state for $\xi = 1$ explains why the energy
behavior as a function of the magnetic field for this carrier,
shown in Fig. \ref{fig:Energ2Cascas} (a), looks like an ordinary
AB spectrum for $B <$ 2 T, where the magnetic field is still not
strong enough to induce the electron confinement in the inner
shell. This is not the case for holes, where both ground and first
excited states are $s-$like.

Figure \ref{fig:2cascaswaves} (b) and (c) show that for smaller
$\xi$, the elliptic inner shell leads to double dot-like states,
as mentioned before in the single shell case, where ground and
first excited states concern to symmetric and anti-symmetric wave
function distributions over the two regions of higher curvature,
for both the electron and the hole. Due to this symmetry issue for
$\xi < 0.96$, the ground state of one carrier is orthogonal to the
first excited state of the other carrier, leading to zero values
of both (1,0) and (0,1) overlaps, which suggests that transitions
between these electron and hole states are forbidden only in the
elliptic case. Besides, due to the fact that the electron and hole
first excited state wave functions exhibit the same symmetry for
$\xi < 0.96$, the (1,1) transitions become allowed (non-zero
overlap) for this range of $\xi$.

\begin{figure}[!h]
\centerline{\includegraphics[width=0.6\linewidth]{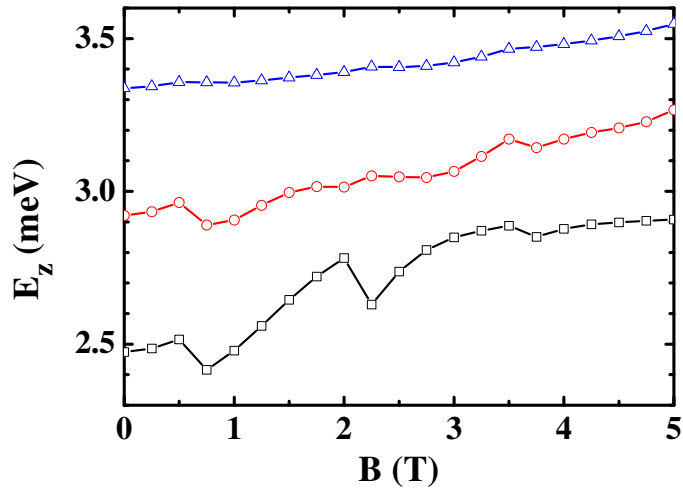}}
\caption{(color online) Exciton binding energies as a function of the magnetic
field for a $R_{1,i}$ = 250 \AA\,, $R_{1,e}$ = 350 \AA\,,
$R_{2,i}$ = 400 \AA\, and $R_{2,e}$ = 500 \AA\,
Al$_{0.3}$Ga$_{0.7}$As/GaAs double shell structure, considering
different values of the inner ring eccentricity $\xi$.}
\label{fig:EbxB2Shells}
\end{figure}

The average radii transitions observed in the Fig.
\ref{fig:Energ2Cascas} clearly affects the ground state binding
energy in this system, specially because they do not occur in the
same way for electrons and holes. Thus, the elliptic distortion of
the inner shell would also help to smooth out the irregularity of
the binding energy as a function of the magnetic field. Indeed, as
one can see in Fig. \ref{fig:EbxB2Shells}, the electron-hole
binding energy $E_z$ responds directly to these average radius jumps, since a different confinement region for each
carrier, represented by different average radii in Fig.
\ref{fig:Energ2Cascas}, leads to a lower binding energy modulus.
These jumps are suppressed as the inner shell becomes elliptic
and, consequently, the abrupt variations in the binding energy are smoothened
out.

Notice that in the single shell case discussed previously,
electrons and holes were already confined in the same region, so
that the exciton naturally had a practically zero net charge and,
consequently, exhibited negligible dependence on the magnetic
field, which was further enhanced as the shell was made elliptic,
due to a slightly different dependence of the carriers confinement
on the eccentricity. The opposite happened for the double shell
structure: the exciton net charge is naturally non-zero and,
consequently, the binding energy is strongly dependent on the
magnetic field, whereas when the inner ring becomes elliptic, it
pushes electrons and holes towards the same confinement region,
creating an almost zero net charge, which suppresses the influence
of the magnetic field on the binding energy.

\section{Conclusions}

We have investigated the influence of an elliptic distortion of
the circular symmetry of core-multishell quantum wires on their
confinement energies and excitonic properties. For a single
confining shell, increasing the eccentricity leads to a
localization of both charge carriers in the points of higher
curvature of the system. Under an in-plane electric field applied
parallel to the direction of distortion, the energy levels behave
as a double dot system, exhibiting energy crossings as a function
of the field intensity. When such a field is applied in different
directions, these crossings are suppressed. This suggests a way to
experimentally probe the eccentricity of a core-multishell wire
and its direction of distortion. For a circular CMS quantum wire,
the electron and hole are both evenly distributed along the
confining shell, creating an effectively neutral exciton which,
therefore, does not interact with an axially applied magnetic
field. On the other hand, electron and hole wavefunctions are not
similarly distributed along the shell in elliptic CMS structures,
creating some regions of effectively non-neutral charge. As a
consequence, the exciton energy exhibits oscillations as a
function of the magnetic field, which is reminiscent of the
Aharonov-Bohm (AB) effect, usually observed in quantum rings.

In the case of two concentric confining shells, we demonstrate
that the barrier between the shells plays an important role on the
definition of the symmetry of the electron and hole
eigenfunctions. In specific cases, it is possible to observe for
one carrier, a ground state wave function that spreads over both
shells and a first excited state with non-zero angular momentum,
whereas for the other carrier, each eigenstates is confined in a
single shell and both have zero angular momentum. This has a
drastic effect not only on the behavior of the energy spectrum of
each carrier under an external magnetic field, but also on the
transition probabilities between electron and hole states.
However, considering an elliptic inner shell, even for
eccentricities as small as $\xi = 0.95$, the electron and hole
states become more similar, as both are confined in the higher
curvature regions. In this case, AB oscillations of the carriers
energies are more clear, and overlaps between their ground states
wave functions are enhanced. Besides, due to the strong
confinement of electrons and holes in these higher curvature
regions, the exciton binding energy becomes less susceptible to
external magnetic fields. Therefore, the inter-shell barrier is
demonstrated to be a useful parameter to tune the allowance of
transitions between electrons and hole states in the circular
case, whereas the eccentricity of the inner ring can be used to
push electron and hole wave functions to the same confinement
regions, smoothening out the dependence of the exciton binding
energy on the external magnetic field.

\ack This work has received financial support from
the Brazilian National Research Council (CNPq), Funda\c{c}\~ao
Cearense de Apoio ao Desenvolvimento Cient\'ifico e Tecnol\'ogico
(Funcap).

\end{document}